\documentclass[prl,twocolumn,showpacs,preprintnumbers,superscriptaddress]
{revtex4}

\usepackage{times}

\usepackage{bm}
\usepackage{graphicx}
\usepackage{amsbsy}
\usepackage{amsmath}
\usepackage{amsfonts}
\usepackage{amsthm}

\begin{document}

\title{Sum Uncertainty Relation in Quantum Theory}
\author{A. K.\ Pati and P. K. Sahu \\
Institute of Physics, Bhubaneswar-751005, Orissa, India}
%\address{Institute of Physics, Bhubaneswar-751005, Orissa, India}
%\affiliation{Institute of Physics, Bhubaneswar-751005, Orissa, India}

%\date{\today}

\def\ra{\rangle}
\def\la{\langle}
\def\ver{\arrowvert}
\begin{abstract}
We prove a new sum uncertainty relation in 
quantum theory which states that the uncertainty in the sum 
of two or more observables is always less than or equal to the sum of 
the uncertainties in corresponding observables. This shows that the quantum 
mechanical uncertainty in any observable is a convex function. 
We prove that if we 
have a finite number $N$ of identically prepared quantum systems, then a joint 
measurement of any observable gives an error $\sqrt N$ less than that 
of the individual measurements. This has application in
quantum metrology that aims to give better precision in the parameter estimation.
Furthermore, 
%using this inequality we prove that speed of a quantum system is
%not an additive quantity.
this proves that a quantum system evolves slowly under the action of a sum Hamiltonian 
than the sum of individuals, even if they are non-commuting.
\end{abstract}


%]

\maketitle

\vskip 2cm

%\begin{multicols}{2}

Unlike in classical physics, there are restrictions in quantum theory on how
accurately one can measure observables even in principle \cite{dirac}. 
This is well 
documented by the famous Heisenberg uncertainty relation for position and 
momentum of a quantum particle \cite{wh}. Later on the uncertainty 
relation was 
generalized for any two non-commuting observables \cite{hpr}. 
It tells that the 
product of uncertainties in two non-commuting observables in a given 
quantum states is greater than or
equal to the average of their commutator in the corresponding quantum state.
The Heisenberg uncertainty relation is then a special case of this generalized 
uncertainty relation.

%However, in the literature there is no sum uncertainty relation for two 
%quantum mechanical observables. 
Here we ask, given two or more observables of
a quantum system if one measures their sum, will the uncertainty 
be more or less
than the sum of their individual uncertainties? It turns out that the error 
introduced in the sum of observables is always less than or equal to the 
sum of the
errors introduced by individual observables. This we term as the 
sum uncertainty relation---which forms the basis to show that quantum 
mechanical 
uncertainty in any observable is actually a convex function. Furthermore, we 
show that if we have a finite number of identically prepared quantum systems, 
then the measurement of 
the collective observable gives an error which is $\sqrt N$ smaller than the
one obtained via individual measurements. We apply these ideas in quantum 
metrology that aims to give better precision in the parameter estimation.
Moreover, we will give some examples and 
illustrate the relation for some simple quantum mechanical systems.
One consequence of the sum uncertainty relation is that a quantum system evolves 
more slowly under the action of a sum Hamiltonian than the sum of either 
separately, i.e., mixing of even non-commuting Hamiltonians slows down the system.

\noindent
{\bf \it Sum uncertainty relation:}
Consider a quantum state $|\Psi\ra$ in a Hilbert space 
${\cal H}$. Let $A$ and $B$ 
are two general observables (they could be commuting or non-commuting) that 
represent some physical quantities. Then, the quantum mechanical uncertainties 
associated with these observables in the state $|\Psi\ra$ are 
defined via $\Delta A^2 =
\la \Psi|A^2|\Psi\ra - \la \Psi|A|\Psi\ra^2$ and $\Delta B^2 =
\la \Psi|B^2|\Psi\ra - \la \Psi|B|\Psi\ra^2$. Similarly, we can define 
the uncertainty in the sum of two observables as 
$\Delta (A+B)^2 =
\la \Psi|(A+B)^2|\Psi\ra - \la \Psi|(A+B)|\Psi\ra^2$.  Here, we address 
the question: what is the relation between  $\Delta (A+B)$, 
$\Delta A$, and $\Delta B$? The following theorem answers this.\\

\noindent
{\bf Theorem:} Quantum fluctuation in  the sum of {\em any} two observables 
is always less than or equal to the sum of their individual fluctuations, i.e.,
$\Delta (A+B) \le \Delta A + \Delta B$. \\

\noindent
{\it Proof:}  Let $A$ and $B$ are two observables which could be commuting or 
non-commuting. Let us define two unnormalized vectors 
$|\Psi_1\ra = (A - \la A\ra)|\Psi\ra$ and $|\Psi_2\ra = 
(B - \la B \ra)|\Psi\ra$, where $\la A\ra = \la \Psi|A|\Psi\ra$ and 
$\la B\ra = \la \Psi|B|\Psi\ra$ are quantum mechanical averages of $A$ 
and $B$, respectively in the state $|\Psi\ra$. Consider the 
norm of sum of two vectors $|\Psi_1 \ra + |\Psi_2\ra$.
This is given by
\begin{eqnarray}
||\Psi_1 + \Psi_2 ||^2 &=& ||\Psi_1 ||^2 + || \Psi_2||^2 + 
2 Re \la \Psi_1|\Psi_2\ra \nonumber\\
&=& \Delta A^2 + \Delta B^2 + 2 Re \la \Psi_1|\Psi_2\ra,
\end{eqnarray}
where $|| \Psi_1||^2 = \la \Psi_1|\Psi_1\ra = 
\la \Psi|(A - \la A\ra)^2|\Psi\ra = \Delta A^2$ and
$|| \Psi_2||^2 =   \la \Psi_2|\Psi_2\ra = \la \Psi|(B - \la B \ra)^2|\Psi\ra 
= \Delta B^2$.
Using the fact that $ Re \la \Psi_1|\Psi_2\ra \le  |\la \Psi_1|\Psi_2\ra|$ and
further using the Schwartz inequality we have $ Re \la \Psi_1|\Psi_2\ra 
\le  ||\Psi_1 ||~ || \Psi_2||$. Then the norm of sum of two vectors satisfy 
\begin{eqnarray}
||\Psi_1 + \Psi_2 ||^2 \le \Delta A^2 + \Delta B^2 + 2 \Delta A \Delta B.
\end{eqnarray}
%then using the triangle inequality we have
%\begin{equation}
%||\Psi_1 + \Psi_2 || \le ||\Psi_1 || + || \Psi_2||,
%\end{equation}
%where $|| \Psi_1||^2 = \la \Psi|(A - \la A\ra)^2|\Psi\ra = \Delta A^2$ and
%$|| \Psi_2||^2 = \la \Psi|(B - \la B \ra)^2|\Psi\ra = \Delta B^2$.
On the other hand direct evaluation of $||\Psi_1 + \Psi_2 ||$ gives 
\begin{eqnarray}
||\Psi_1 + \Psi_2 ||^2 & =& \la \Psi| [(A -\la A\ra) + 
(B - \la B\ra)]^2 \Psi\ra \nonumber\\
%&=& \la A^2 \ra - \la A\ra^2 + \la B^2 \ra - \la B\ra^2 \nonumber\\
%& + & \la AB + BA\ra - 2 \la A\ra \la B\ra \\
&=& \la \Psi|(A +B)^2|\Psi\ra -  \la \Psi|(A +B)|\Psi\ra^2 \nonumber\\
&=& \Delta (A+B)^2.
\end{eqnarray}
Thus, (2) and (3) imply that 
\begin{equation}
 \Delta (A+B) \le  \Delta A + \Delta B
\end{equation}
which is the sum uncertainty relation. Hence, the proof.

The physical meaning of the sum uncertainty relation is that if we have 
an ensemble of quantum systems then the ignorance in totality is 
always less than the sum of the individual ignorance. In case of two 
observables, if we prepare a 
large number of quantum systems in the state $|\Psi\ra$, and then perform 
the measurement of $A$ on some of those systems and 
%perform the measurement of 
$B$ on some others, then the standard deviations in $A$ plus $B$ will be 
more than the standard deviation in the measurement of $(A+B)$
on those systems. Hence, it is always advisable to go for 
`total measurement' if we want to minimize the error.

In fact, it is not difficult to see that if we have more than two 
observables (say three observables $A$, $B$, and $C$), 
then the sum uncertainty relation will read as
\begin{equation}
 \Delta (A+B+C) \le  \Delta A + \Delta B + \Delta C.
\end{equation}
In general for $N$ observables $A_1, A_2, \cdots, A_N$, we will have the sum 
uncertainty relation as
\begin{equation}
 \Delta (\sum_i A_i) \le  \sum_i \Delta A_i, (i=1,2,\cdots, N).
\end{equation}

\noindent
{\it Convexity of quantum uncertainty:}
The above inequality brings out an important property of the quantum 
uncertainty with convexity of a function.
To be specific, we will show that 
%It may be worth mentioning 
the quantum mechanical uncertainty in any observable is actually 
a convex function. Recall that $f$ is a convex function if 
\begin{equation}
 f(\sum_i p_i x_i) \le  \sum_i p_i f(x_i), (i=1,2,\cdots, N)
\end{equation}
where $p_i$'s satisfy $0 < p_i < 1$, $\sum_i p_i =1$  and $x_i$ is in 
the set $S$ \cite{marcus}.
Note that under scaling transformation of an operator 
$A \rightarrow \lambda A$, the quantum mechanical uncertainty transforms as
$\Delta A \rightarrow \lambda \Delta A$. Using this fact we can have a more 
general sum uncertainty relation. For example, if we have a sum observable 
$\sum_i p_i A_i$, with all $p_i$'s
as positive numbers, then we have the following sum uncertainty relation
\begin{equation}
 \Delta (\sum_i p_i A_i) \le  \sum_i p_i \Delta A_i, (i=1,2,\cdots, N).
\end{equation}
The meaning of this general sum uncertainty relation is that `mixing of 
commuting or non-commuting operators' always decreases the uncertainty.

Now, note that 
in (8) if the positive numbers $p_i$'s satisfy $0 < p_i < 1$ and 
$\sum_i p_i =1$, then $\Delta$ is indeed a convex 
function. Furthermore, it is known that if $f_1, f_2, \ldots f_n$ are convex 
functions on ${\bf R}$ and $p_i \ge 0$, $(i=1,2, \ldots n)$, then 
$f(x) = \sum_i p_i f_i(x)$ is also a convex function on ${\bf R}$. 
By using this property of convex function, we can draw the following 
conclusion.
Suppose we have several quantum states $|\Psi_1\ra, |\Psi_2\ra, \ldots 
|\Psi_n\ra$ and the quantum mechanical uncertainties in $A$, in the above 
states are $\Delta_1, \Delta_2, \ldots \Delta_n$. Then, it 
follows that $\Delta = \sum_i p_i \Delta_i$ is a convex function.
The usefulness of convexity of quantum uncertainty may be similar to the 
entropy of a quantum system. 
It may be mentioned that entropy of a quantum state is a concave function.
Also, we know that entropy signifies the information content of a quantum 
state. So, in that sense one can think of negative of quantum uncertainty 
as a concave function and it may represent as a `measure of information' 
(ignorance). Thus, this property unravels another feature of quantum 
mechanical uncertainty.

\noindent
{\it Error in collective and individual measurements:}
We can test the inequality (6) explicitly by considering an ensemble 
that consists of 
$N$-identically prepared quantum systems. Let each system be in the state
$|\Psi\ra$. Therefore, the combined state vector of $N$-particle is given by
\begin{equation}
|\Psi\ra^{\otimes N} = |\Psi\ra_1 \otimes |\Psi\ra_2 \otimes \cdots 
\otimes |\Psi\ra_N. 
\end{equation}

Let us first measure an observable $A$ on each particle individually 
(not collectively). The individual observables of interest are 
$A_1 = A \otimes I \otimes  \cdots I$,  $A_2 = 
I \otimes A \otimes  \cdots I$,  .... and $A_N = 
I \otimes I \otimes  \cdots A$. Then, one can check that the average 
of $A_i (i=1,2, \ldots N) $ 
in the state $|\Psi \ra^{\otimes N} = \la \Psi|A|\Psi\ra$. Similarly, 
the uncertainty in each $A_i$ is $\Delta A$. Therefore, the sum of 
uncertainties in the individual measurements is $\sum_i \Delta A_i = N \Delta A$.

Now, suppose we perform measurement of the sum observable on $N$-copies. 
The sum observables is given by
\begin{eqnarray}
A_S = \sum_i A_i &=& A \otimes I \otimes  \cdots I + 
I \otimes A \otimes  \cdots I \nonumber\\
&+& \cdots + 
I \otimes I \otimes  \cdots A.
\end{eqnarray}
The quantum mechanical uncertainty in the observable $A_S$ in the state 
$|\Psi\ra^{\otimes N}$ is given by
\begin{equation}
\Delta A_S^2 = {^{\otimes N}}\la \Psi | A_S^2|\Psi\ra^{\otimes N} 
- (^{\otimes N}\la \Psi | A_S|\Psi\ra^{\otimes N})^2 .
\end{equation}
Note that ${^{\otimes N}}\la \Psi | A_S^2|\Psi\ra^{\otimes N} = 
N \la \Psi|A^2|\Psi\ra + 2 N  \la \Psi|A|\Psi\ra^2$ and 
$^{\otimes N}\la \Psi | A_S|\Psi\ra^{\otimes N} = N \la \Psi|A|\Psi\ra$.
Hence, $\Delta A_S^2 = N \Delta A^2$. This implies that the quantum mechanical
uncertainty in the sum observable is  
$\Delta A_S = \sqrt{ N} \Delta A$. Therefore, the sum uncertainty relation 
reads as
\begin{equation}
\sqrt{ N} \Delta A \le N \Delta A
\end{equation}
which is clearly satisfied. This analysis also suggests that the error in the
total measurement goes as $\sqrt{ N} \Delta A$, whereas the sum of errors in 
the individual 
measurement goes as $N \Delta A$. Thus, there is an overall 
$\sqrt{ N}$ improvement in the error of measurement of sum observable 
with $N$-copies.

\noindent
{\it Parameter estimation and quantum metrology:}
Precision measurement which requires estimation of some parameter to its 
highest accuracy is an important problem. If one uses laws of quantum theory,  
then in the measurement of some parameter the precession can be enhanced. 
This a 
topic of great study in quantum metrology. In this scheme one prepares a probe
state $|\psi_0\ra$, applies a unitary operator $U(\theta)$ that depends on the 
parameter $\theta$ to be estimated and then measures some observable $X$ on 
the resulting state $|\psi(\theta)\ra$. If $U(\theta) = \exp(-i\theta H)$ 
where $H$ is a Hermitian operator, then using the Mandelstam-Tamm 
uncertainty relation \cite{mt,bc}, we have 
\begin{eqnarray}
\Delta X \Delta H \ge \frac{1}{2} \big| \partial \la X \ra/\partial
\theta \big|.
\end{eqnarray}
The precision with which one can estimate $\theta$ is given by
\begin{eqnarray}
\delta \theta = 
\Delta X /\big| \partial \la X \ra/\partial \theta \big| \ge 
\frac{1}{2 \Delta H},
\end{eqnarray}
where  $\Delta X$, $\Delta H$, and $\la X \ra$ have their usual meaning in 
the quantum state $|\psi(\theta)\ra$.
Therefore, if we want to minimize the error in estimating the parameter, 
we have 
to minimize $\Delta X$ or maximize $\Delta H$. How to achieve that goal is 
the subject of quantum metrology \cite{vlm}. 
It turns out that using quantum entangled 
probe states or entangling unitary operator one can achieve better and better 
precision in the parameter estimation.

Recently, Giovannetti {\it et al} \cite{seth} have shown that 
that using entangled probe state one can achieve an 
enhancement that scales as $1/N$. More recently, it was shown by 
Roy and Braunstein \cite{rb} 
that if one exploits entangling unitary operator then one obtains 
an exponential enhancement 
in the parameter estimation. In particular, by choosing an 
appropriate Hamiltonian  one can apply the unitary operator 
$U=e^{-i\theta H}$ and generate a $N$-qubit 
state given by (for details see \cite{rb})
\begin{eqnarray}
|\psi_H(\theta) &=& e^{-i\theta H}|00 \ldots 00\ra \\
&=&  \cos (2^{N-1} \theta)|00\ldots 00\ra - i \sin (2^{N-1} \theta)
|11\ldots 11\ra. \nonumber
\end{eqnarray}
Then, by measuring the observable $X= \otimes_{i=1}^{N} P_j$, 
where $P_j =|0\ra_j\la 0|$ one can estimate $\theta$ as given 
by $\delta \theta = 1/2^N$ which is the exponential
enhancement in the precision.

Here, we show that there are other class of measurement strategies also 
which can give the same precision. 
Suppose, instead of measuring the product observable we measure the sum 
observable, i.e.,
measure $X = \sum_i P_i = P_S=
P_1 \otimes I \otimes  \cdots  I + 
I \otimes P_2 \otimes  \cdots \otimes I + \cdots + 
I \otimes I \otimes  \cdots \otimes P_N$. 
Then, the 
precision in the parameter estimation is given by 
\begin{eqnarray}
\delta \theta = \Delta P_S /\big| \partial \la P_S \ra/\partial \theta \big|.
\end{eqnarray}
The quantum uncertainty and average for $P_S$ in the state 
$|\psi_H(\theta)\ra$ are given by
\begin{eqnarray}
\Delta P_S = 
%N \cos (2^{N-1} \theta) \sin (2^{N-1} \theta), 
\frac{N}{2} \sin (2^N \theta),~
\la P_S \ra = 
N \cos^2 (2^{N-1} \theta).
\end{eqnarray}
Therefore, the precision in the estimation of the parameter $\theta$ is 
$1/2^N$. Similarly, if we measure individually these projectors then the 
corresponding precision in the parameter $\theta$ will be 
\begin{eqnarray}
\delta \theta = \sum_i \Delta P_i /\big| \partial (\sum_i \la P_i \ra)/
\partial \theta \big|.
\end{eqnarray}
One can check that for the state $|\psi_H(\theta) \ra$, we have  
$\sum_i \Delta P_i = \frac{N}{2} \sin (2^N \theta) $ and 
$\sum_i \la P_i \ra = N \cos^2 (2^{N-1} \theta)$.
Thus, again we see that $\delta \theta = 1/2^N$. Hence, for the entangling 
unitary, joint measurement and individual measurements give the same precision 
as obtained in \cite{rb}. The physical explanation is now clear. 
This is happening because, the sum uncertainty relation is 
saturated for these observables and the equality holds.

One may wonder if by increasing the resources and by 
exploiting the sum uncertainty relation (i.e. the idea that the measurement of 
sum observable minimizes the error leading to better precision) one can 
enhance the precision in the parameter better than the exponential \cite{rb}. 
However, as we will see this is not the case.
Let us imagine that we have an ensemble of $N$-probe
states. The number of copies of $N$-probe states are finite (say) $M$. On each 
of the $N$-probe state we apply the entangling unitary operator as suggested by
Roy and Braunstein \cite{rb}. But there is no further interaction between 
these 
$M$-copies. Then, the combined state of $MN$ probe state is given by
\begin{equation}
|\psi_H(\theta) \ra^{\otimes M} = |\psi_H(\theta)\ra_1 \otimes 
|\psi_H(\theta) \ra_2 \otimes \cdots 
\otimes |\psi_H(\theta) \ra_M . 
\end{equation}
On these collection of $MN$ probe states we measure a sum observable. The 
observable of interest is 
\begin{eqnarray}
\Pi_S  &=& \Pi_1 \otimes I_2 \otimes  \cdots  I_M + 
I_1 \otimes \Pi_2 \otimes  \cdots \otimes I_M + \cdots   \nonumber\\ 
&+& I_1 \otimes I_2 \otimes  \cdots \otimes \Pi_M,
\end{eqnarray}
where $\Pi_j, (j=1,2, \ldots M)$ are product of projection operators on 
$j$th copy of the $N$-probe state. To be clear, 
$\Pi_1 = P= \otimes_{i=1}^{N} P_i$ on the $1$st 
$N$-probe state,  $\Pi_2 = P= \otimes_{i=1}^{N} P_i$ on the $2$nd 
$N$-probe state and so on. The precision in the measurement of the sum 
observable $\Pi_S$ is given by 
\begin{eqnarray}
\delta \theta = \Delta \Pi_S /\big| \partial \la \Pi_S \ra/
\partial \theta \big|.
\end{eqnarray}
One can check that the quantum uncertainty and average in $\Pi_S$ 
for the $MN$ probe state $|\psi_H(\theta) \ra^{\otimes M} $ are given by
\begin{eqnarray}
\Delta \Pi_S = \frac{\sqrt{M}}{2} \sin (2^N \theta), 
\la \Pi_S \ra = M \cos^2 (2^{N-1} \theta).
\end{eqnarray}
Therefore, the precision in the estimation of the parameter 
$\theta$ is given by
\begin{eqnarray}
\delta \theta = 1/\sqrt{M} 2^N.
\end{eqnarray}
This result apparently may give an impression that 
this is better than exponential \cite{rb}. But, if we re-express the result in terms of actual 
resource used, i.e., the number $K=MN$, then the precision $\delta \theta = 
\sqrt N 2^{K(\frac{M-1}{M})} \frac{1}{\sqrt K 2^K}$, which is lower than the exponential.
Therefore, it is always not the case that by using more resources one can enhance
the precision.

%{\it Examples and illustrations:}
Now, we give few further applications of the sum uncertainty relation 
in quantum theory. First, we apply to Hamiltonian systems and second we apply
to the speed of quantum mechanical systems. 

\noindent
{\it Uncertainty in the Hamiltonian:}
One immediate application is that for any quantum mechanical system, 
the total Hamiltonian $H$ consists of kinetic and potential energy, i.e., 
$H = T+V $ and using the sum uncertainty relation we have 
$\Delta H \le \Delta T + \Delta V $. Thus, the uncertainty in the total 
energy in any state is bounded by the sum of uncertainties in the kinetic
and potential energy. This result is interesting, in the sense that if we
want to do energy measurement with minimal error, then do not measure kinetic 
and potential energy separately.  Always measure the total energy 
because the quantum mechanical uncertainty is less in that case. 
Also, this shows that the uncertainty in
the total energy is less than the uncertainty in the kinetic energy 
in the position basis. However in the 
momentum basis, the uncertainty in
the total energy is less than the uncertainty in the potential energy. These 
observations may have many applications in the complex quantum systems.

\noindent
{\it Sub-additivity of quantum speed:}
%Another counter intuitive application is the following. 
%Classically, suppose we apply a force $F_1$ to a particle, then it will move with 
%a speed $v_1$. Similarly, under a force $F_2$, the particle will move with a 
%speed $v_2$. Suppose, we add these forces, i.e., now the particle moves 
%under the applied force $F= F_1 + F_2$. 
%Using Newton's law one can see that the speed with which the particle will move
%in the space is given by $v=v_1 + v_2$.  
%Now, let us analyze the situation quantum mechanically. 
%However, we will show that the speed with which a quantum system 
Here, we ask whether the speed of evolution of a state vector through Hilbert 
space behaves like the classical speed. 
In what follows, we will show that the speed with which a quantum system 
evolves under two Hamiltonians (commuting or non-commuting) are not added up. 
%Note that the `speed' now is not in the ordinary space but in the quantum 
%state space or in the projective Hilbert space. 
(Note that classically, if a particle is subjected to two force fields, 
then the speed of a particle is added up.)

In quantum theory, when a system evolves under some Hamiltonian $H$, 
then the state
undergoes a continuous time evolution, i.e., $|\Psi(0)\ra \rightarrow 
|\Psi(t)\ra = \exp(-iHt)|\Psi(0)\ra$. One can ask how fast does the system 
evolve in time.
Then, the rate at which it evolves is nothing but the speed of 
transportation of the 
state vector in the projective Hilbert space \cite{aa,akp}. This is defined as 
%\begin{equation}
$v = \frac{dD}{dt}$,
%\end{equation}
where $dD$ is the infinitesimal distance between nearby quantum states 
$|\Psi(t)\ra$ and $|\Psi(t+dt)\ra$. The distance function is given by 
\begin{equation}
dD^2 = (1 - |\la \Psi(t)|\Psi(t+dt)\ra|^2) = \frac{dt^2}{\hbar^2} \Delta H^2,
\end{equation}
where $\Delta H$ is the usual uncertainty in the Hamiltonian in the state 
$|\Psi\ra$.
%Thus, the speed is given by $v= \Delta H/\hbar$. 
Therefore, the speed at which a quantum system evolves is nothing but the 
uncertainty in the Hamiltonian of the system, i.e.,$v= \Delta H/\hbar$. 
This is the geometric meaning of quantum fluctuation: more the fluctuation 
in the Hamiltonian, faster the system will evolve. 

Now, imagine that a quantum system evolves under a Hamiltonian $H_1$, then the 
speed is given by $v_1= \Delta H_1/\hbar$. 
Similarly, if this evolves under a Hamiltonian $H_2$, then the speed 
is given by $v_2= \Delta H_2/\hbar$. 
Suppose, now the system evolves under the Hamiltonian $H = H_1 + H_2$. What 
will be the speed? Will the total speed be $v = v_1 + v_2$? 
The answer is no. Using the 
sum uncertainty relation we see that $v = \Delta H/\hbar \le 
\Delta H_1/\hbar + \Delta H_2/\hbar$. In other words, 
the quantum speed obeys the relation 
\begin{equation}
v \le v_1 + v_2.
\end{equation}
The meaning of this equation is that in general a quantum system will evolve 
more slowly under the action of a sum Hamiltonian than the sum of either separately. 
This is a non-trivial result, in the sense that this holds for generic 
Hamiltonians be they commuting or non-commuting. 
This is something counter intuitive which arise due to 
quantum mechanical nature of the associated observables and also the fact that 
quantum systems obey the Schr{\"o}dinger equation and not the Newton equation.
%This behavior is simply not there in the classical world.
Also, it may be noted that if we have a Hamiltonian $H = H_1 - H_2$, the speed
will obey the relation 
\begin{equation}
v \le v_1 + v_2.
\end{equation}
This is due to the fact that $\Delta (-A) = \Delta A$, i.e., quantum 
mechanical uncertainty is an even function.
%However, classically if a particle moves under a force $F_1- F_2$, then the 
%speed will be $v = v_1 - v_2$.  This is another interesting feature of 
%quantum speed which is not seen in the classical world.

One may ask whether the velocity operator of a quantum system obeys the sub-additivity
condition. We will show that in general this may not. But the average of the velocity 
operator may obey a kind of sub-additivity condition. Note that using the Heisenberg 
equation of motion, the velocity operator can be defined as $v = 
\frac{1}{i\hbar}[x, H]$, where
$x$ is the position and $H$ is the Hamiltonian. Then, the magnitude of the average of
the velocity operator will obey $|\la v \ra| \le \frac{2}{\hbar} \Delta x \Delta H$. 
Now, if we have a Hamiltonian $H_1$, then the velocity operator will obey  
$|\la v_1 \ra| \le \frac{2}{\hbar} \Delta x \Delta H_1$. Similarly, for 
a Hamiltonian $H_2$, the velocity   
$|\la v_2 \ra| \le \frac{2}{\hbar} \Delta x \Delta H_2$. This implies that 
$|\la v_1 \ra|_{\rm max} = \frac{2}{\hbar} \Delta x \Delta H_1$ and 
$|\la v_2 \ra|_{\rm max} = \frac{2}{\hbar} \Delta x \Delta H_2$. Now, 
using the sum uncertainty relation we have 
\begin{equation}
|\la v \ra| \le \frac{2}{\hbar} \Delta x (\Delta H_1 + \Delta H_2).
\end{equation}
This suggests that $|\la v \ra| \le |\la v_1 \ra|_{\rm max} + |\la v_2 \ra|_{\rm max}$.
This is another interesting application of the sum uncertainty relation.

\noindent
{\it Conclusion:} We have proved a new sum uncertainty relation for general
 observables in quantum theory which shows that quantum mechanical uncertainty
in the sum of two or more observables is always less than or equal to the sum 
of quantum  uncertainties in the individual observables. We have also proved 
that the quantum mechanical uncertainty is indeed a convex function. This property 
suggests that there is some analogy between quantum uncertainty and entropy of 
a quantum mechanical system.
We have shown that if we have a finite number of identically prepared 
quantum states, 
%then the error in the sum measurement of any observable goes as 
%$\sqrt{ N} \Delta A$ whereas the sum of errors in the individual measurement 
%goes as $N \Delta A$. This shows that there is an overall $\sqrt{ N}$ 
then there is an overall $\sqrt{ N}$ 
improvement in the error of 
measurement of the sum observable with $N$-copies. As an important application 
we have explained why the measurement of the 
sum and individual observables can give the same exponential precision. 
Also, we have shown that using more resources one cannot have a precision 
better than the exponential one. 
%We have discussed few interesting applications of this new inequality. 
In addition, we prove that in general a quantum system evolves 
more slowly under the action of a sum Hamiltonian than the sum of either 
separately. It is expected that the sum uncertainty relation will have wider 
applications in a variety of context like quantum computation, quantum information 
theory and many body quantum systems. 

\renewcommand{\baselinestretch}{1}

\noindent
{\bf Acknowledgment:} We are thankful to S. L. Braunstein for useful remarks.

%\end{multicols}

\end{document}